\documentclass[conference]{IEEEtran}
\IEEEoverridecommandlockouts
\usepackage{cite}
\usepackage{amsmath,amssymb,amsfonts}
\usepackage{algorithmic}
\usepackage{graphicx}
\usepackage{subfig}
\usepackage{booktabs}
\usepackage{caption}
\usepackage{textcomp}
\usepackage{multirow}
\usepackage{xcolor}
\def\BibTeX{{\rm B\kern-.05em{\sc i\kern-.025em b}\kern-.08em
    T\kern-.1667em\lower.7ex\hbox{E}\kern-.125emX}}
\begin{document}

\title{A GAN-Based Image Transformation Scheme for Privacy-Preserving Deep Neural Networks
\thanks{This work was partially supported by Grant-in-Aid for Scientific
Research(B), No.17H03267, from the Japan Society for the Promotion Science.}
}

\author{\IEEEauthorblockN{Warit Sirichotedumrong and Hitoshi Kiya}
\IEEEauthorblockA{\textit{Department of Computer Science, Tokyo Metropolitan University, 
Hino-shi, Tokyo, Japan}}}

\maketitle

\begin{abstract}
We propose a novel image transformation scheme using generative adversarial networks (GANs) for privacy-preserving deep neural networks (DNNs). 
The proposed scheme enables us not only to apply images without visual information to DNNs, but also to enhance robustness against ciphertext-only attacks (COAs) including DNN-based attacks. 
In this paper, the proposed transformation scheme is demonstrated to be able to protect visual information on plain images, and the visually-protected images are directly applied to DNNs for privacy-preserving image classification. 
Since the proposed scheme utilizes GANs, there is no need to manage encryption keys. 
In an image classification experiment, we evaluate the effectiveness of the proposed scheme in terms of classification accuracy and robustness against COAs.
\end{abstract}


\section{Introduction}
\label{sec:intro}
Deep neural networks (DNNs) have become one of the most powerful approaches for 
 solving complex problems for many applications\cite{lecun2015deep,Donahue2014,Krizhevsky2012}, 
 such as for computer vision, biomedical systems, natural language processing, and speech recognition.
DNNs gain an advantage by using a large amount of data to extract representations of relevant features, 
so the performance is significantly improved\cite{michael2018on}. 
However, DNNs have been deployed in security-critical applications, such as facial recognition, biometric authentication, and medical image analysis.

Instead of using local servers, it is very popular for data owners to use cloud servers to compute a huge amount of data, 
but several data security issues have to be considered for utilizing DNNs, 
such as data privacy, data leakage, and unauthorized data access. 
Moreover, DNNs can be vulnerable if adversaries carry out model inversion attacks\cite{fredrikson2015model,Basu2019arxiv} to obtain trained data from the trained model. 
Hence, privacy-preserving DNNs have become an urgent challenge.
In this paper, we focus on protecting visual information of images before uploading the images to cloud environments.

Various perceptual image encryption methods have been proposed to protect visual information of images\cite{kurihara2015pcs,Gaata2016,warit2018icme,tanaka2018iccetw,itier2019tcsvt,chuman2019tifs,warit2019apsipa_trans,warit2019icip,warit2019access}.
In contrast to information theory-based encryption (like RSA and AES), images encrypted by the perceptual encryption methods can be directly applied to some image processing algorithms.

A few previous works\cite{tanaka2018iccetw,warit2019icip,warit2019access} have been proposed not only to protect the visual information of images but also to be applicable to DNNs. 
However, the visual information of images encrypted by using these methods have been pointed out to be reconstructed if adversaries carry out DNN-based attacks\cite{warit2019gcce,warit2019access,gan_attack}.
 
This paper proposes a novel image transformation scheme for privacy-preserving DNNs that enables us not only to apply images without visual information to DNNs, but also to enhance robustness against COAs. 
Since the proposed scheme utilizes the framework of generative adversarial networks (GANs), there is no need to manage encryption keys. 
In an image classification experiment, the proposed scheme is demonstrated to enable us not only to maintain high classification accuracy as that of conventional methods but also to have stronger robustness against COAs including DNN-based attacks than conventional ones.  

\begin{figure}[!t]
\captionsetup[subfigure]{justification=centering}
\centering
\subfloat[]{\includegraphics[clip, height=2.8cm]{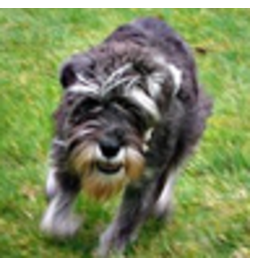}
\label{fig:label-A}}
\hfil
\subfloat[]{\includegraphics[clip, height=2.8cm]{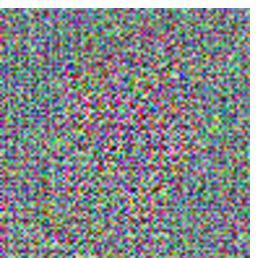}
\label{fig:label-B}}
\\
\subfloat[]{\includegraphics[clip, height=2.8cm]{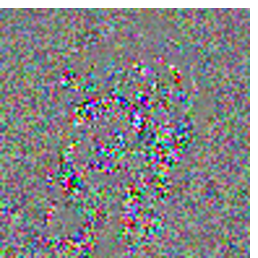}
\label{fig:label-C}}
\hfil
\subfloat[]{\includegraphics[clip, height=2.8cm]{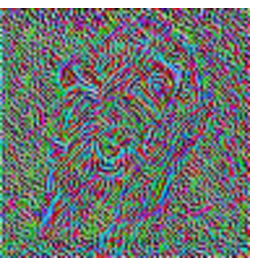}
\label{fig:label-D}}
\caption{Example of encrypted images. (a) Original image ($U \times V = 96 \times 96$). (b) Tanaka's scheme\cite{tanaka2018iccetw} (c) Pixel-based encryption\cite{warit2019icip,warit2019access}. 
(d) Proposed scheme.}
\label{fig:eximages}
\end{figure}

\section{Related Work}
\label{sec:related}
\subsection{Image Encryption for Privacy-Preserving DNNs}
\label{ssec:visual}
Various perceptual image encryption methods have been proposed to protect visual information of plain images\cite{kurihara2015pcs,Gaata2016,warit2018icme,tanaka2018iccetw,itier2019tcsvt,chuman2019tifs,warit2019apsipa_trans,warit2019icip,warit2019access}.
Visually-protected images generated by using an encryption method consist of  pixels, as shown in Figs.\,\ref{fig:eximages}(b) and \,\ref{fig:eximages}(c). 
Therefore, the encrypted images can be directly applied to image processing algorithms. 
Some of encryption methods have been studied for applying encrypted images to traditional machine learning algorithms, such as
support vector machine (SVM), under the use of the kernel trick\cite{maekawa2018apsipa,Takahiro2019ieice}
, but they cannot be applied to DNNs\cite{warit2019access}.

There are three encryption methods\cite{tanaka2018iccetw,warit2019icip,warit2019access} for privacy-preserving DNNs.
The first method is Tanaka's scheme\cite{tanaka2018iccetw}, which utilizes an adaptation network prior to DNNs to reduce the influence of image encryption.
The second\cite{warit2019icip} is a pixel-based method, which enables data augmentation in the encrypted domain, under the use of a common encryption key to every image. 
The third one\cite{warit2019access} enables us to assign an independent encryption key to every image; therefore, there is no need to manage keys.
However, these methods do not consider robustness against DNN-based attacks\cite{warit2019gcce}. 
Hence, this paper aims to produce visually-protected images having robustness against various COAs including DNN-based ones, while maintaining high classification accuracy.

\begin{figure}[t]
\centering
\includegraphics[width =7.5cm]{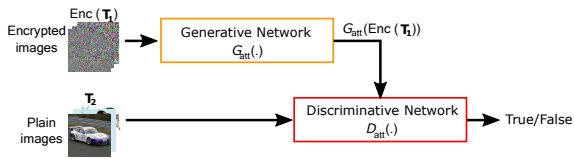}
\caption{The framework of GAN-based attack (GA)}
\label{fig:dcgan}
\vspace{-0.5cm}
\end{figure}

\subsection{Robustness against Ciphertext-only Attacks}
In this paper, we focus on robustness against COAs, such as brute-force attacks and DNN-based attacks. 
Conventional encryption methods\cite{tanaka2018iccetw,warit2019icip,warit2019access} have large key spaces, so they are robust against brute-force attacks. 
However, they are pointed out not to have enough robustness against DNN-based attacks\cite{warit2019access}.

There are two type of DNN-based attacks: unshared weight attack (UWA)\cite{warit2019gcce} and GAN-based attack (GA)\cite{gan_attack}.
Although the pixel-based encryption with independent encryption keys\cite{warit2019access} has stronger robustness against UWA than that with a common key, 
robustness against GA has not been evaluated yet. 
The detail of GA is described in this section.

\begin{figure}[t]
\centering
\includegraphics[width =6.5cm]{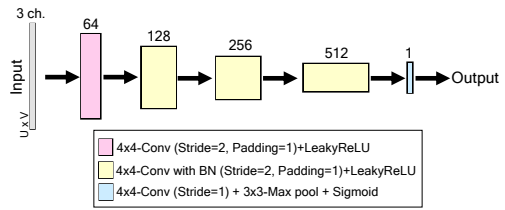}
\caption{Discriminative network $D_{att}(.)$}
\label{fig:discriminator}
\vspace{-0.5cm}
\end{figure}

\begin{figure*}[t]
\centering
\includegraphics[width =15.5cm]{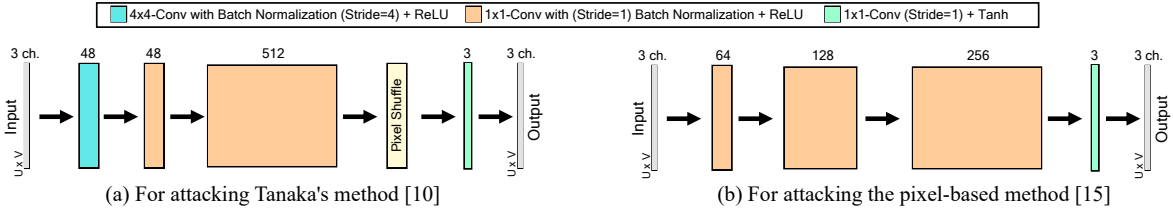}
\caption{Architectures of generative network $G_{att}(.)$.}
\label{fig:generator}
\vspace{-0.5cm}
\end{figure*}

Figure\,\ref{fig:dcgan} illustrates the framework of GA which consists of one generative network and one discriminative network. 
The generative network ($G_{att}(.)$) reconstructs visual information on plain images from the encrypted images while the discriminative network ($D_{att}(.)$) distinguishes the difference between reconstructed images and the plain ones. 
The network architectures of $D_{att}(.)$ and $G_{att}(.)$ used in this paper are illustrated in Fig.\,\ref{fig:discriminator} and Fig.\,\ref{fig:generator}, respectively. 

To train GA, a set of training images $\mathbf{T}$ is equally divided into two image sets: $\mathbf{T_1}$, and $\mathbf{T_2}$, and then $\mathbf{T_1}$ is encrypted by using an encryption scheme ($Enc(.)$).
Eventually, $Enc(\mathbf{T_1})$ and $\mathbf{T_2}$ are employed for training $G_{att}(.)$ and $D_{att}(.)$ (See Fig.\,\ref{fig:dcgan}).


\begin{figure}[t]
\centering
\subfloat[Model training and testing]{\includegraphics[clip, width =6.5cm]{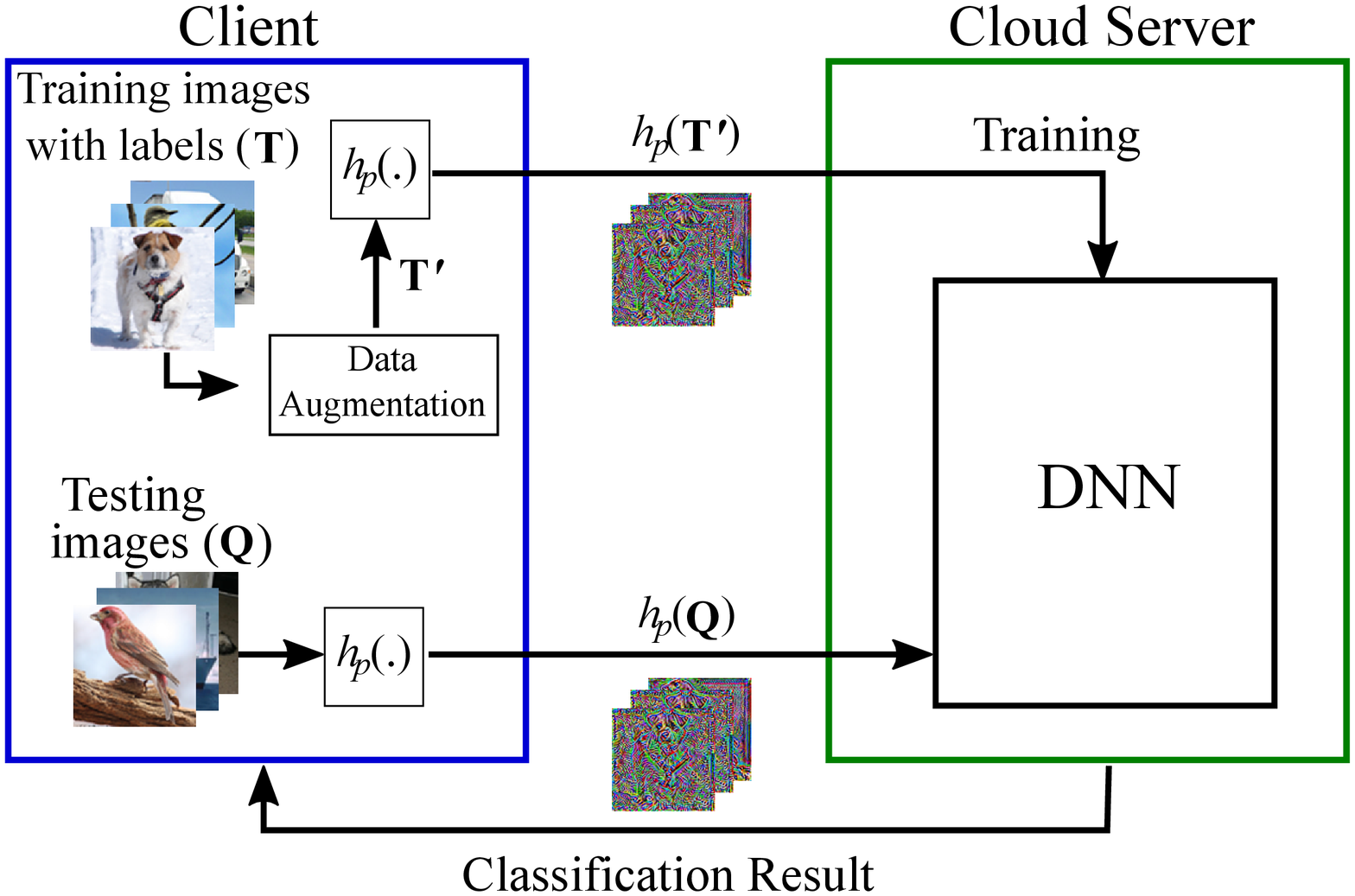}
\label{}}
\\
\subfloat[Design of transformation network $h_p(.)$]{\includegraphics[clip, width =7.5cm]{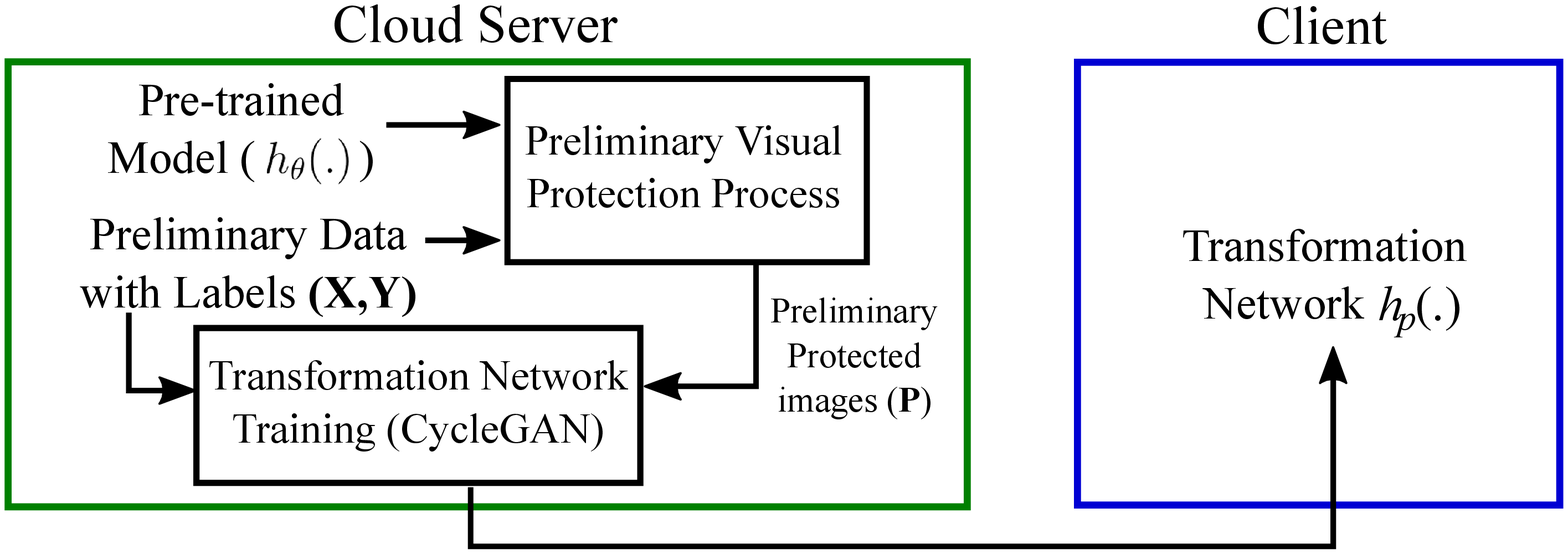}
\label{}}
\caption{Proposed scheme for privacy-preserving DNNs.}
\label{fig:system}
\vspace{-0.2cm}
\end{figure}

\section{Proposed Scheme}
\label{sec:proposed}
\subsection{Overview}
Figure\,\ref{fig:system}(a) illustrates the use of a transformation network ($h_p(.)$) for protecting training and testing images in DNNs.
In order to obtain $h_p(.)$, as depicted in Fig.\,\ref{fig:system}(b), the proposed scheme has two processes: preliminary visual protection and transformation network training. 
In the preliminary visual protection process, a set of preliminary protected images ($\mathbf{P}$) is generated from a pre-trained model ($h_{\theta}(.)$) by using a set of training images ($\mathbf{X}$) and the corresponding labels ($\mathbf{Y}$). 
$\mathbf{X}$ with $\mathbf{Y}$ , and $\mathbf{P}$ are utilized for training $h_p(.)$ which is used for generating a set of visually-protected images ($\mathbf{X_p}$).
Namely, $\mathbf{X_p}=h_p(\mathbf{X})$.  

Hence, images used for training and testing DNNs are protected by using $h_p(.)$, as shown in Fig.\,\ref{fig:system}(a).

\subsection{Preliminary Visual Protection Process}
\label{ssec:pvop}
We aim to generate a protected image ($x_p$) from a plain image ($x$) with a label ($y$) by adding some perceptible noise $\Delta$; therefore, $x_p=x+\Delta$.
Let $x_i$ and $y_i$ denote $i$-$th$ image with a label, where $x_i\in\mathbf{X}$ and $y_i\in\mathbf{Y}$.
In contrast to adversarial examples\cite{pgd}, protected images are generated by minimizing the loss with respect to input $x$ with $\Delta$ as given by

\begin{equation}
\underset{\Delta\in\mathcal{P}}{minimize}\frac{1}{m}\sum_{i=1}^{m}\mathcal{L}(h_\theta(x_i+\Delta),y_i),
\end{equation}
where $\mathcal{P}$ is a perturbation set, and $\mathcal{L}(.)$ is a loss function. 
The perturbation set $\mathcal{P}$, which is used for protecting visual information on plain images, can be written as
\begin{equation}
\mathcal{P}=\{\Delta:{\|\Delta\|}_{\infty}\leq\epsilon\},
\end{equation}
where $\epsilon$ is the magnitude of the perturbation. The proposed method is inspired by adversarial attacks called projected gradient descent (PGD). We minimize the loss, but the adversarial attack maximizes it.
At the $(t+1)^{th}$ iteration, the protected image (${x}_p^{t+1}$) is
\begin{equation}
\label{eq:pga}
{x}_p^{t+1}=\prod_{x+\Delta}({x}_p^t+{\alpha}sign(\nabla_x(-\mathcal{L}(\theta,x,y)))),
\end{equation}
where $\alpha$ is a step size. In this paper, $\Delta$ is iteratively added to an image for 50 iterations.

Therefore, $h_\theta(.)$ is expected to precisely classify the protected images, although the images have no visual information.

\subsection{Transformation Network}

To obtain a transformation network $h_p(.)$, we utilize unpaired image-to-image translation using cycle-consistent adversarial networks (CycleGAN)\cite{zhu2017unpaired} to convert 
an image from the plain domain to the visually-protected domain. 
\subsubsection{Formulation}
\begin{figure}[t]
\centering
\includegraphics[width =6cm]{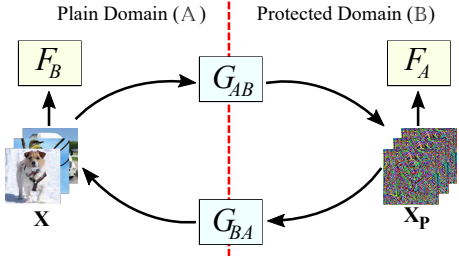}
\caption{CycleGAN architecture}
\label{fig:cyclegan}
\vspace{-0.5cm}
\end{figure} 
Let $\mathbb{A}$ and $\mathbb{B}$ denote the plain domain and the visually-protected domain, respectively.
As shown in Fig.\,\ref{fig:cyclegan}, CycleGAN consists of two generative networks: $G_{AB}: \mathbb{A}\rightarrow\mathbb{B}$, and $G_{BA}: \mathbb{B}\rightarrow\mathbb{A}$, which transform an image between two domains. 
In addition, there are two discriminative networks $F_A$ and $F_B$. 
$F_A$ discriminates the difference between $x$ and $G_{BA}(x_p)$ while $F_B$ distinguishes the difference between $x_p$ and $G_{AB}(x)$. 
The objective of training CycleGAN is to obtain $G_{AB}$ which transforms plain images to visually-protected images, namely, $h_p(.)=G_{AB}$.

\subsubsection{Loss Function}
According to \cite{zhu2017unpaired}, the CycleGAN utilizes two loss functions: adversarial loss ($\mathcal{L}_{ad}$), and cycle consistency loss ($\mathcal{L}_{cyc}$). 
Therefore, the full objective function ($\mathcal{L}_{gan}$) can be expressed as
\begin{equation}
\begin{split}
\mathcal{L}_{gan}(G_{AB},G_{BA},F_A,F_B)=&\lambda\mathcal{L}_{ad}(G_{BA},F_A,x,x_p)\\
&+\lambda\mathcal{L}_{ad}(G_{AB},F_B,x_p,x)\\
&+\mathcal{L}_{cyc}(G_{BA},G_{AB}),
\end{split}
\end{equation}
where $\lambda$ is utilized for weighting the relation between $\mathcal{L}_{ad}$ and $\mathcal{L}_{cyc}$.

In order to ensure that images generated by $h_p(.)$ have no visual information while maintaining high classification accuracy, we propose a new cycle consistency loss which can be written as

\begin{equation}
\label{eq:cyc}
\mathcal{L}_{cyc}={\gamma_1}\cdot\mathcal{L}_{p}+{\gamma_2}\cdot\mathcal{L}_{c}+{\gamma_3}\cdot\mathcal{L}_{r},
\end{equation}
where $\mathcal{L}_{p}$, $\mathcal{L}_{c}$, and $\mathcal{L}_{r}$ denote a perceptual loss, classification loss, and pixel-wise reconstruction loss, respectively.

We first aim to maximize the perceptual loss ($\mathcal{L}_p$)\cite{Johnson2016Perceptual} to protect the visual information. 
Since the CycleGAN consists of two cycles: $\mathbb{A}\rightarrow\mathbb{B}$, and $\mathbb{B}\rightarrow\mathbb{A}$.
The perceptual loss of our CycleGAN, $\mathcal{L}_{p}$, can be expressed as
\begin{equation}
\label{eq:feat}
\mathcal{L}_p=\mathcal{L}_{feat}(x,G_{AB}(x))+\mathcal{L}_{feat}(x,G_{BA}(x_p)).
\end{equation}
In accordance with \cite{Johnson2016Perceptual}, the feature loss ($\mathcal{L}_{feat}$) is calculated by
\begin{equation}
\mathcal{L}_{feat}(x_i,{\hat{x}_i})=\frac{1}{{C_k}{H_k}{W_k}}{\|{\phi}_k({\hat{x}}_i)-{\phi}_k(x_i)\|}_{2}^{2},
\end{equation}
where ${\phi}_k(x_i)$ is a feature map with a size of ${C_k}\times{H_k}\times{W_k}$, obtained by using the $k^{th}$ layer of a network when image $x_i$ is fed \cite{Johnson2016Perceptual}.
Note that $\hat{x}_i$ is a transformed image generated by $G_{AB}(.)$ and $G_{BA}(.)$. 
In this paper, we use VGG16\cite{Simonyan15vgg}, which is pre-trained with ImageNet, as a network for ${\phi}_k(x_i)$.
Note that we utilize the feature maps of the second ReLU fuction of VGG16.

Moreover, to maintain high classification accuracy, we solve the minimization problem of the classification loss ($\mathcal{L}_{cls}$) by utilizing $h_{\theta}$ from Section\,\ref{ssec:pvop}, where $\mathcal{L}_{cls}$ is given by the cross entropy loss. 
The classification loss of the CycleGAN ($\mathcal{L}_c$) is
\begin{equation}
\label{eq:cls}
\mathcal{L}_c=\mathcal{L}_{cls}(h_\theta(G_{AB}(x),y))+\mathcal{L}_{cls}(h_\theta(G_{BA}(x_p),y)
\end{equation}

As in \cite{zhu2017unpaired}, we also utilize the following reconstruction loss ($\mathcal{L}_r$). 
\begin{equation}
\label{eq:rec}
\mathcal{L}_r={\|{G_{BA}(G_{AB}(x))-x}\|}_2^2+{\|{G_{AB}(G_{BA}(x_p))-x_p}\|}_2^2.
\end{equation}

In an experiment, we employ the architecture of U-Net\cite{unet} for $G_{AB}$ and $G_{BA}$, and the architectures of $F_A$ and $F_B$ are the same as in \cite{zhu2017unpaired}
\section{Experiments}
\label{sec:experiments}

Image classification experiments were carried out to confirm the effectiveness of the propose scheme.

\subsection{Dataset}

We employed two image datasets: CIFAR-10 and CIFAR-100, for training and evaluating the proposed scheme.
The first one is CIFAR-10 dataset which contains $32 \times 32$-pixel color images and consists of 50K training images\cite{cifar10}.
Namely, the transformation network $h_p(.)$ was trained with CIFAR-10 dataset.

To evaluate the classification performance, we employed CIFAR-10 and CIFAR-100\cite{cifar10}, which contain $32 \times 32$-pixels color images and consists of 50K training images and 10K test images in 10 classes and 100 classes, respectively. 
We utilized CIFAR-100, to prove that the proposed scheme is applicable to any datasets, even when the model was trained by using CIFAR-10.

For training and testing two DNN-based attacks: UWA\cite{warit2019gcce} and GA\cite{gan_attack}, we employed STL-10 dataset\cite{pmlr-v15-coates11a}, which contains $96 \times 96$-pixel color images and consists of 5K training images and 8K testing images.

\subsection{Training Transformation Network}

VGG13\cite{Simonyan15vgg} with batch normalization was trained by using CIFAR-10 so that this model was considered as $h_\theta(.)$.
Then, we obtained preliminary protected images ($\mathbf{P}$) for training the CycleGAN with the proposed loss function in {\eqref{eq:cyc}}.

CycleGAN was trained for 5000 epochs by using the Adam optimizer\cite{kingma2014adam} with a learning rate of 0.0002, and a momentum parameters of ${\beta}_1=0.5$ and ${\beta}_2=0.999$, where $\lambda=0.4$, ${\gamma_1}=-1$, ${\gamma_2}=0.4$, and ${\gamma_3}=0.9$.
As a result, $h_p(.)$ was obtained and used for generating visually-protected images. An example of images generated by $h_p(.)$ is illustrated in Fig.\,\ref{fig:eximages}(d), where Fig.\,\ref{fig:eximages}(a) is the original one.

\begin{table}
\centering
\caption{Image classification accuracy and average structural similarity (SSIM) values of reconstructed images by DNN-based attacks. (N/A: not available)}
\label{tbl:result}
\begin{tabular}{cc|cc|cc}
\toprule
\multicolumn{2}{c|}{\multirow{3}{*}{\shortstack{Encryption\\ Scheme}}}&\multicolumn{2}{c|}{SSIM values}&\multicolumn{2}{c}{Accuracy (\%)}\\
\cmidrule{3-6}
&&UWA&GA&CIFAR&CIFAR\\
&&\cite{warit2019gcce}&\cite{gan_attack}&-10&-100\\
\midrule
\multicolumn{2}{c|}{Proposed Scheme}&N/A&\textbf{0.0956}&90.73&67.36\\
\midrule
Pixel-based&Same&0.1715&0.2688&91.29&67.17\\
\cmidrule{2-6}
\cite{warit2019icip,warit2019access}&Different&0.0425&0.1527&91.14&67.28\\
\midrule
\multicolumn{2}{c|}{Tanaka's Scheme\cite{tanaka2018iccetw}}&N/A&0.7152&88.96&68.13\\
\midrule
\multicolumn{2}{c|}{Plain images}&N/A&N/A&95.05&77.24\\
\bottomrule
\end{tabular}
\vspace{-0.3cm}
\end{table}
 
\subsection{Robustness against Attacks}

\subsubsection{Training Conditions}
GA was trained for 100 epochs by using the Adam optimizer\cite{kingma2014adam} with a learning rate of 0.0002, a momentum parameter of ${\beta}=0.5$, and a batch size of 64. 
For reconstructing images encrypted by Tanaka's scheme and the pixel-based method\cite{warit2019icip,warit2019access}, we employed the network architectures in Figs.\,\ref{fig:generator}(a) and \,\ref{fig:generator}(b), respectively.
U-Net\cite{unet} was utilized as $G_{att}(.)$ in order to reconstruct transformed images generated by the proposed scheme.

In contrast, UWA\cite{warit2019gcce} was trained for 70 epochs by using stochastic gradient descent (SGD) with momentum for 70 epochs. 
The initial learning rate was set to 0.1, and was lowered by a factor of 10 at 40 and 60 epochs. We utilized a weight decay of 0.0005, a momentum of 0.9, and a batch size of 128. 

In all experiments, we did not carry out any data augmentation and pre-processing to training images.

\subsubsection{Results}
\begin{figure}[!t]
\captionsetup[subfigure]{justification=centering}
\centering
\subfloat[Tanaka's scheme]{\includegraphics[clip, height=2.8cm]{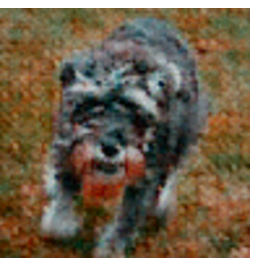}
\label{fig:label-A}}
\hfil
\subfloat[Pixel-based (Same Key)]{\includegraphics[clip, height=2.8cm]{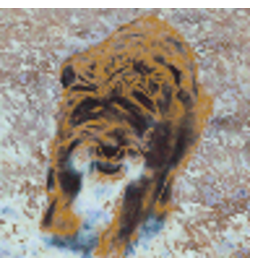}
\label{fig:label-B}}
\\
\subfloat[Pixel-based (Different Key)]{\includegraphics[clip, height=2.8cm]{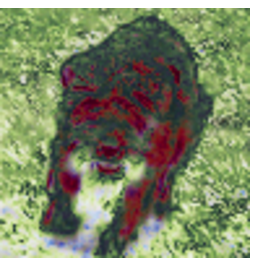}
\label{fig:label-C}}
\hfil
\subfloat[The proposed scheme]{\includegraphics[clip, height=2.8cm]{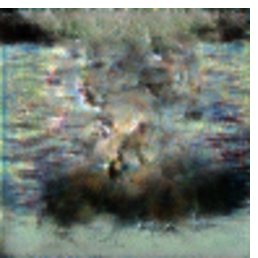}
\label{fig:label-D}}
\caption{Example of reconstructed images.}
\label{fig:recon}
\vspace{-0.3cm}
\end{figure}

Examples of reconstructed images are shown in Fig.\,\ref{fig:recon}, where Fig.\ref{fig:eximages}(a) is the original one. 
Although visual information of images encrypted by the conventional methods was reconstructed by GA, 
it was difficult to reconstruct the visual information on plain images from images encrypted by the proposed scheme.

In addition, we measured average structural similarity (SSIM) values between original images and reconstructed ones, where lower values mean lower visual information. 
As shown in Table\,\ref{tbl:result}, the proposed scheme provided the lowest SSIM values in encryption methods, namely, images generated by the proposed scheme have higher robustness against GA than conventional ones. 

\subsection{Image Classification}

\subsubsection{Training Conditions}
We used deep residual networks (ResNet)\cite{resnet} without pre-trained weights for evaluating the effectiveness of the proposed scheme in terms of image classification accuracy. 

We evaluated the performance by using ResNet with 18 layers (ResNet-18).
The models with ResNet were trained by using stochastic gradient descent (SGD) with momentum for 200 epochs. 
The initial learning rate was set to 0.1, and was lowered by a factor of 5 at 60, 120, 160 epochs. We set a weight decay and momentum to 0.0005 and 0.9.

As shown in Fig.\,\ref{fig:system}(a), the models were trained with live augmentation (random cropping with padding = 4 and random horizontal flip). 
A client firstly carried out standard normalization and data augmentation to generate pre-processed training data ($\mathbf{T'}$) from $\mathbf{T}$. 
Eventually, the client protected $\mathbf{T'}$ by using $h_p(.)$ and then uploaded $h_p(\mathbf{T'})$ to the servers to train DNN models.
As a result, the DNN models were trained by 50K images for both CIFAR-10 and CIFAR-100.
\subsubsection{Results}
We compared the proposed scheme with the pixel-based image encryption\cite{warit2019access} in terms of classification accuracy and robustness against COAs. 
As shown in Table.\,\ref{tbl:result}, the proposed method provided almost the same classification accuracy as the pixel-based method. 
Therefore, the proposed scheme enables us not only to maintain the classification accuracy as that of the pixel-based method, but also to enhance robustness against DNN-based attacks.

\section{Conslusion}
\label{sec:conclusion}

We presented a novel image transformation scheme using generative adversarial networks (GANs) for privacy-preserving deep neural networks (DNNs). 
The proposed scheme enables us not only to apply images without visual information to DNNs, but also to enhance robustness against ciphertext-only attack (COAs). 
In this paper, the proposed scheme was demonstrated to be able to protect visual information on plain images, and the visually-protected images were directly applied to DNNs for privacy-preserving image classification. 
Since the proposed scheme utilizes GANs, there is no need to manage encryption keys. 
In an experiment, we evaluated the effectiveness of the proposed scheme in terms of classification accuracy and robustness against attacks. 
The experimental results confirmed that the proposed scheme enables us not only to maintain high classification accuracy as that with the pixel-based method, but also to enhance robustness against DNN-based attacks.

\begin{small}
\bibliographystyle{IEEEtran}

\end{small}
\end{document}